\documentclass[a4paper,11pt]{article}
\pdfoutput=1 
\usepackage{jcappub} 
\usepackage[T1]{fontenc} 
\usepackage[utf8]{inputenc}
\usepackage[english]{babel}
\usepackage{graphicx} 
\usepackage{placeins}
\usepackage{subcaption}
\usepackage{mathrsfs}
\usepackage[dvipsnames]{xcolor}
\usepackage{enumerate}

\newcommand{\LG}{\mathcal{L}}
\newcommand{\port}{\sqcap}
\newcommand{\pd}{\partial}
\newcommand{\med}{\phi}
\newcommand{\dm}{S}
\newcommand{\gkinp}[2]{{#1^{#2}}_{\mu\nu}\widetilde{{#1_{#2}}}^{\mu\nu}}

\title{\textcolor{RoyalPurple}{Di-photon excess at LHC and the gamma ray excess at the Galactic Centre}}

\author[a,1]{Andi Hektor,\note{Corresponding author.}}
\author[a,b]{and Luca Marzola}

\affiliation[a]{National Institute of Chemical Physics and Biophysics; R\"{a}vala pst. 10,\\ 10143 Tallinn, Estonia}
\affiliation[b]{Institute of Physics, University of Tartu; Ravila 14c, 50411 Tartu, Estonia}

\emailAdd{andi.hektor@cern.ch}
\emailAdd{luca.marzola@ut.ee}

\abstract{
Motivated by the recent indications for a 750~GeV resonance in the di-photon final state at the LHC, in this work we analyse the compatibility of the excess with the broad photon excess detected at the Galactic Centre. Intriguingly, by analysing the parameter space of an effective models where a 750~GeV pseudoscalar particles mediates the interaction between the Standard Model and a scalar dark sector, we prove the compatibility of the two signals. We show, however, that the LHC mono-jet searches and the Fermi LAT measurements strongly limit the viable parameter space. We comment on the possible impact of cosmic antiproton flux measurement by the AMS-02 experiment.
}
\begin{document}
\maketitle 
\flushbottom

\section{Introduction}
\label{sec:intro}

The ATLAS and CMS experiments at CERN's Large Hadron Collider (LHC) have recently reported an excess in the di-photon channel peaked at a centre-of-mass energy of about 750~GeV~\cite{diphotonATLAS, diphotonCMS}. The significance of the signal detected by CMS is 2.6$\sigma$, while the ATLAS collaboration measured a 3.6$\sigma$ excess. As the two indications seem to be compatible within the resolutions of the detectors, the di-photon excess could be a first manifestation of the physics beyond the Standard Model (SM) at the LHC~\cite{Harigaya:2015ezk, Backovic:2015fnp, Pilaftsis:2015ycr, DiChiara:2015vdm, Nakai:2015ptz, Buttazzo:2015txu, Knapen:2015dap, Angelescu:2015uiz, Mambrini:2015wyu, Franceschini:2015kwy}. In this light, for a di-photon final state, the Landau-Yang theorem asserts that the speculated 750~GeV particle must either have spin 0 or 2. Whereas future analysis of the photon angular distribution could discriminate between these hypotheses, the new physics interpretation is already being challenged by the absence of signals in channels~\cite{diphotonATLAS, diphotonCMS} as the di-jet~\cite{Khachatryan:2015dcf} or the $t\bar t$~\cite{Khachatryan:2015sma} one, as well as the di-boson and the di-lepton one, which are complementary in traditional SM extensions like the minimal supersymmetric or the two-Higgs doublets model. Another puzzling aspect of the reported excess is the apparently large width inferred from the signal. This suggests that the di-photon anomaly could be induced by new particles having almost degenerate masses or, alternatively, by a single particle with a large invisible decay width. The latter possibility is certainly fascinating as the speculated 750 GeV particle could be the missing link between the visible and the dark sector of our Universe.

Dark matter (DM) accounts for about the 26\% of the energy density of the present Universe and is usually modelled in a relic abundance of weakly-interacting massive particles (WIMPs). This scheme is motivated by the ``WIMP miracle'': particles with masses and annihilation cross sections set by the electroweak scale automatically yield a DM abundance in the ballpark of the measured value through the freeze-out mechanism (for reviews see~\cite{Jungman:1995df, Bertone:2004pz}). The WIMP paradigm is backed by traditional SM extensions, such as the supersymmetric models, and motivates the ongoing dedicated searches that are based on three complementary approaches. Firstly, the direct detection experiments investigate the potential elastic scattering between DM and Standard Model particles, which supposedly proceed through a weak interaction. Secondly, collider experiments aim to detect the production of DM by investigating the missing transverse energy, for instance, in mono-jet and mono-photon events. Finally, the fact that DM particles could annihilate in regions of space characterised by a high DM abundance like the centres of galaxies or galaxy clusters, provides the basis of DM indirect detection experiments. These aim to distinguish the DM annihilation traces from the astrophysical background. In this regard, the Galactic Centre (GC) and the dwarf satellite galaxies of the Milky Way should provide the best hypothetical signal/background ratio.

Interestingly, the Fermi LAT telescope discovered in 2009 a spatially extended $\gamma$-ray excess at the GC region in the energy range of 1-5~GeV~\cite{Atwood:2009ez}. Despite many following studies confirmed the detected anomaly~\cite{Goodenough:2009gk, Hooper:2010mq, Abazajian:2010zy, Boyarsky:2010dr, Hooper:2011ti, Abazajian:2012pn, Gordon:2013vta, Macias:2013vya, Abazajian:2014fta, Daylan:2014rsa, Lacroix:2014eea, Calore:2014nla}, which in the present paper we refer to as the Galactic Centre Excess (GCE), we remind the reader that the GC is an extremely complex environment populated with stars, stellar relics, dust, gas, cosmic rays and the central black hole. Then, on one hand, isolating a clean signature of potential DM annihilations from the astrophysical background is extremely challenging and, for instance, it was even found that millisecond pulsars~\cite{Yuan:2014rca} or ultra-energetic events form the past~\cite{Petrovic:2014uda} could offer a competing explanation of GCE. On the other hand, it is also true that the estimates of the DM annihilation cross section implied by the GCE and the morphology of the same signal match the expected DM thermal freeze-out cross-section and Galactic density profile.

Explaining the GCE within the WIMP paradigm requires a rather light DM candidate. For instance, if DM annihilates prevalently to a $b \bar b$ final state, the best fit of the signal is obtained for $m_{\rm DM} = 48.7^{+6.4}_{-5.2}$~GeV. Lower values $m_{\rm DM} = 9.96^{+1.05}_{-0.91}$~\cite{Calore:2014nla} are obtained by fitting the GCE with the $\tau^-\tau^+$ channel. Unfortunately, such a light WIMP candidate is affected by many constraints ranging from the $\gamma$-ray searches of Fermi LAT~\cite{Ackermann:2015tah,Ackermann:2015zua} to the precise CMB measurements~\cite{Galli:2009zc, Slatyer:2009yq, Huetsi:2009ex, Cirelli:2009bb, Hutsi:2011vx, Evoli:2012qh, Madhavacheril:2013cna, Ade:2015xua}. On top of that, the strong bounds cast by the direct detection experiments XENON100~\cite{Aprile:2012nq} and LUX~\cite{Akerib:2013tjd} strongly disfavour the weak-scale values of the elastic scattering cross section supported by the WIMP paradigm. The mentioned DM searches then seem to indicate alternative models of DM characterised by relatively large annihilation cross sections, explaining the detected GCE signal, but possessing a suppressed DM-nucleon interaction. Model of this kind naturally arise once the dark sector and the visible sector of the Universe are put into contact by a pseudoscalar mediator, as in ~\cite{Boehm:2014hva} for instance.

Motivated by the di-photon excess at the LHC and by the photon excess at the Galactic Centre, in this paper we consider a scenario in which a 750~GeV pseudoscalar particle reproduces the former via effective couplings with gluons and photons. The width of the signal is addressed by the invisible decay of the pseudoscalar mediator to a scalar DM candidate, stabilised at this level by effect of a $Z_2$ symmetry proper of the dark sector. In this setup, we show that beside accommodating the LHC di-photon excess while respecting the constraints from complementary channels, our effective model successfully reproduces the measured DM abundance and fits, as well, GCE. Having that the production of the 750 GeV mediator at the LHC can proceed via gluon fusion, we use the DM annihilation into gluons to reproduce the GCE spectrum~\cite{Calore:2014nla}.

Whereas aspects of this work have already been individually presented in literature \cite{Huang:2015svl, Backovic:2015fnp, Bi:2015uqd, Mambrini:2015wyu}, we propose here a first organic study of the mentioned scenario posing particular emphasis on the impact of the $\gamma$-ray line constraints brought by the Fermi LAT measurements~\cite{Ackermann:2015lka}, of the mono-jet searches at the LHC \cite{Khachatryan:2014rra,Aad:2015zva} and of the possible future cosmic antiproton measurements~\cite{AMS02:antiprotons}. As we will show below, although the model is able to reproduce the mentioned excesses in a vast part of its parameter space, the mono-jet searches and the Fermi LAT measurements strongly limit the scenario. The paper is organised as follows: in the next section we detail the considered model, while in section~\ref{sec:constraints} we show our analysis and quantify the impact of the relevant constraints. We summarise our results in section~\ref{sec:Conclusions}.

\section{The model}
\label{sec:model}

In order to explain both the di-photon excess and the GCE, we extend the SM content by introducing a pseudoscalar particle $\med$ and a scalar DM candidate $\dm$. The corresponding free-field Lagrangian is  
\begin{equation}
	\label{eq:lg0}
	\LG_0 
	= 
	\frac{1}{2} 
	\left[(\pd\med)^2 + m_\med^2 + (\pd \dm)^2 + m_\dm^2 \right] 
\end{equation} 
where we set $m_\med = 750$ GeV as required by the LHC signal. The mass of our DM candidate will be determined by the analysis proposed in the next section.

The contact with the SM is provided by the effective portal Lagrangian
\begin{equation}
	\LG_\port
	=
	\frac{c_1}{v} \, \med \, \gkinp{B}{}
	+
	\frac{c_2}{v} \, \med \, \gkinp{W}{a}
	+
	\frac{c_3}{v} \, \med \, \gkinp{G}{a}
\end{equation}
where we normalised the coefficients $c_i$, $i\in\{1,2,3\}$, to the Higgs vacuum expectation value $v=246$ GeV and defined the dual field-strength tensors according to $\widetilde{F}_{\mu\nu}=\frac{1}{2}\epsilon_{\mu\nu\alpha\beta}F^{\alpha\beta}$ for $F\in\{B, W, G\}$. 

The interaction of the psudoscalar mediator with DM is instead regulated by the parity violating Lagrangian
\begin{equation}
	\LG_{\dm\med}
	=
	\frac{1}{2} \, g_S \, \med \, \dm^2
\end{equation}
in a way that the total Lagrangian that we consider is simply $\LG = \LG_{SM} + \LG_0 + \LG_\port + \LG_{\dm\med}$.

In the following we assume that the pseudoscalar mediator is produced at the LHC via gluon-gluon fusion, yielding at the resonance a cross section for the di-photon production of~\cite{Franceschini:2015kwy}
\begin{equation}
	\sigma(pp\to\phi\to\gamma\gamma) 
	= 
	\frac{ C_{gg} \, \Gamma_{gg}\, \Gamma_{\gamma\gamma} }{m_\med\,s\,\Gamma_\med}
\end{equation}
where $C_{gg}=2137$ at $\sqrt s = 13$ TeV and we neglected a sub-dominant quark contribution. The relevant partial decay widths of the pseudoscalar mediator are given by
\begin{align}
	\Gamma_{\gamma\gamma} 
	& = 
	\frac{\left(c_1 c_W^2 + c_2 s_W^2\right)^2}{4\pi}\frac{m_\med^3}{v^2}
	\\
	\Gamma_{gg}
	& =
	\frac{2 c_3^2}{\pi}\frac{m_\med^3}{v^2}
\end{align}
whereas in  the total width $\Gamma_\med$ we accounted also for the contributions of the Electroweak gauge bosons and DM. In particular, for the latter
\begin{equation}
	\Gamma_{DM} = \frac{g_S^2 \, \beta_S}{32\pi\,m_\med}
\end{equation} 
being $\beta_S = \sqrt{1-4m_\dm^2/m_\med^2}$ and $c_W$, $s_W$ respectively the cosine and sine of the Weinberg angle.
\section{Compatibility of the two excesses}
\label{sec:constraints}

\begin{figure}[h]
\centering
\includegraphics[width=.85\textwidth]{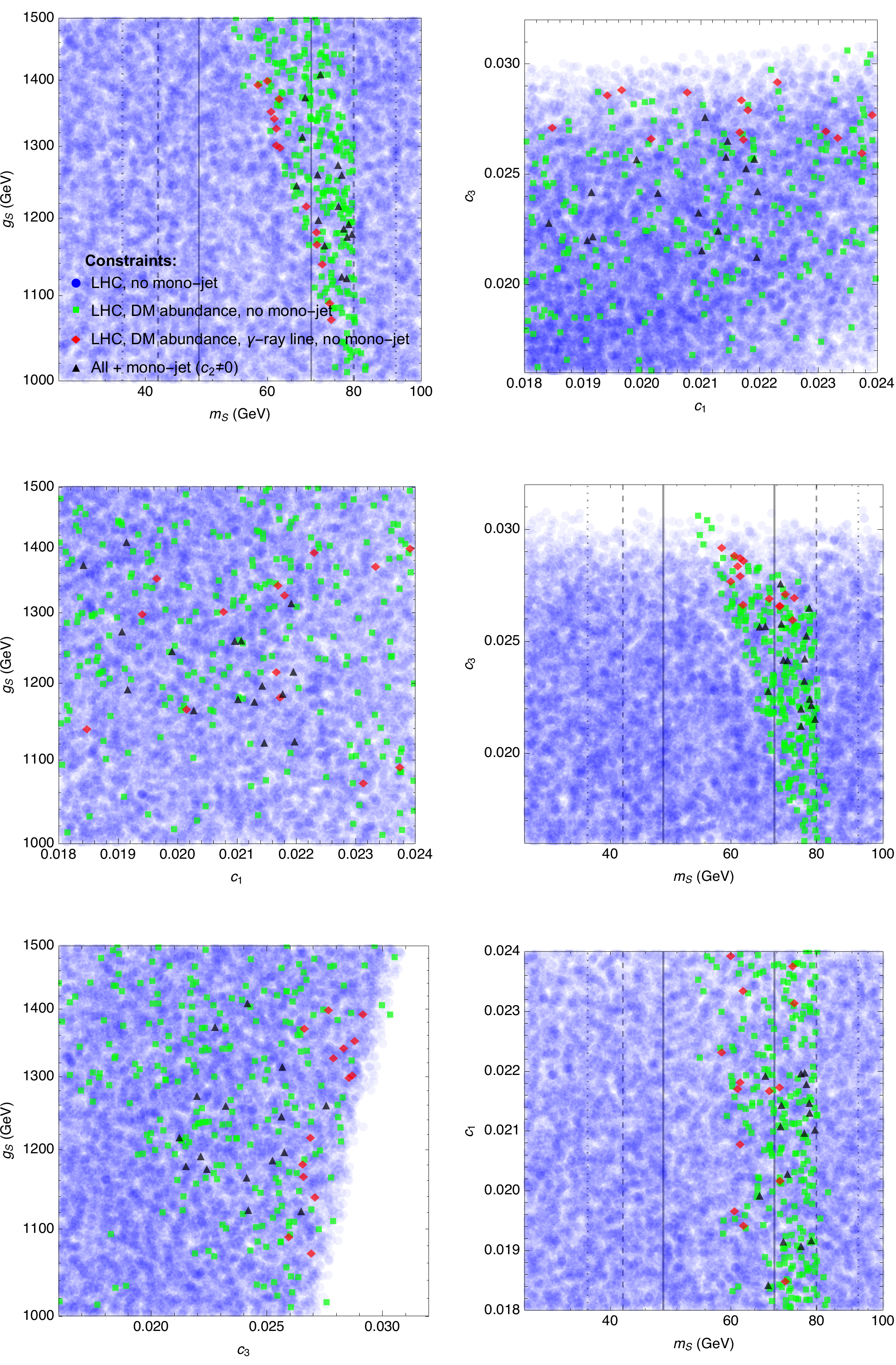}
\caption{Scatter plots for the parameters of the model $c_1$, $c_3$, $g_S$, $m_S$ ($c_2$ is not shown to save space). The blue, green and red points satisfy all the LHC constraints on the 750 GeV resonance but the mono-jet one. Both the green and red points yield a correct DM abundance but only the latter satisfy also the $\gamma$-ray line constraint from Fermi LAT. The points marked as black triangles satisfy also the mono-jet bound on top of all the mentioned constraints. The gray vertical lines indicate the 1, 2 and 3$\sigma$ regions for the Galactic Centre Excess fit via the $SS \to gg$ channel~\cite{Calore:2014nla}.}\label{fig:scatterplots}
\end{figure}

Adopting the setup illustrated in the previous section, we now scan the parameter space of the model attempting to reproduce both the LHC di-photon excess and the GCE. We explored a region in the parameter space bounded by
\begin{align}
	\label{eq:scan_mS}
	& m_S = [30, 100] \text{ GeV} \\
	\label{eq:scan_gS}
	& g_S = [10^3, 1.8 \times 10^3] \text{ GeV} \\
	\label{eq:scan_c1}
	&  c_1 = [0.018, 0.024] \\
	\label{eq:scan_c2}
	& c_2 = [-0.11, -0.075] \\
	\label{eq:scan_c3}
	& c_3 = [0.016, 0.032]
\end{align}
generating a total of $10^5$ random points. The range of $m_S$ is a priori motivated by the fits of GCS via the $SS \to gg$ channel, for instance~\cite{Calore:2014nla}. 
We show in Fig.~\ref{fig:scatterplots} projections on two-parameter subspaces of the considered parameter space. We progressively apply the following constraints: (i) the LHC bounds, (ii) the cosmological abundance of DM as measured by the Planck collaboration~\cite{Ade:2015xua} and (iii) the $\gamma$-ray constraints reported by the Fermi LAT~\cite{Ackermann:2015lka} experiments. Finally we test the points of our parameter space against the GCE, assuming for the $\gamma$-ray line constraint an average DM  velocity of $v_{\rm Gal} = 10^{-3}$. 
\subsection{LHC constraints}
Starting with the LHC bounds, we implemented the model in \textsc{MadGraph5\_aMC@NLO}~\cite{Alwall:2014hca} and computed the relevant cross sections by using the \textsc{Cteq6l1} parton distribution functions~\cite{Pumplin:2002vw}. In order to reproduce the signal we require that
\begin{equation}
	 5\text{ fb} \lesssim \sigma(p p \to \phi \to \gamma \gamma)\lesssim 15 \text{ fb}
\end{equation}
at $\sqrt{s} = 13$ TeV, imposing as well that the total width of the pseudoscalar mediator do not exceed the bound
\begin{equation}
	\Gamma_\med \lesssim 75 \text{ GeV}.
\end{equation}
On top of that, in order to comply with the negative results of searches in the mentioned complementary channels, we consider the following limits resulting from the LHC run-I \cite{CMS:2014onr,Khachatryan:2015cwa,Aad:2015kna,Aad:2015mna,Aad:2014aqa,Aad:2015agg}:
\begin{align}
	\label{eq:lhcb4}
	& \sigma(pp \to \phi \to jj) < 1 \text{ pb} \\
	\label{eq:lhcb1}
	& \sigma(pp \to \phi \to ZZ) < 12 \text{ fb} \\
	\label{eq:lhcb2}
	& \sigma(pp \to \phi \to W^+W^-) < 40 \text{ fb} \\
	\label{eq:lhcb3}
	& \sigma(pp \to \phi \to Z\gamma) < 4 \text{ fb}
\end{align}
valid for $\sqrt s = 8$ TeV. The blue dots in the following figures show the points of the parameter space satisfying these LHC constraints.

On top of that, we analysed the impact of the LHC mono-jet searches, which yield a conservative bound of \cite{Barducci:2015gtd,Khachatryan:2014rra,Aad:2015zva,Bi:2015uqd}
\begin{equation}
	\sigma(p p \to jSS) < 8 \text{ pb}
\end{equation}
at $\sqrt{s} = 13$ TeV. The points of the parameter space that, on top of all the considered constraints, satisfy also the mono-jet bound are represented by black triangles in the plots below.

\FloatBarrier
\subsection{Cosmological abundance and direct detection constraints}
We address DM phenomenology by computing the freeze-out relic abundance of our candidate $S$:
\begin{equation}
	\Omega_{DM}h^2 \simeq \frac{ 1.04 \, x_f}{M_{P}\sqrt{g_\star(x_f)}(a+3b/x_f)} \times 10^9 \text{ GeV}^{-1}
\end{equation}
where $M_P$ is the Planck mass and $g_\star$ the effective number of relativistic degrees of freedom, computed here at the freeze-out epoch. In our scenario, the DM freeze-out proceeds mainly through the annihilation of $S$ into gluons, mediated by the pseudoscalar $\phi$. The relevant averaged cross section has been expanded to the first order in the velocity squared yielding the coefficients
\begin{align}
	& a = \frac{16 \, c_3^2 \, g_S^2 \, m_{\dm}^2}
	{\pi v^2\left[\left(m_\med^2-4m_\dm^2\right)^2+m^2_\med\,\Gamma_\med^2\right]}\\
	& b = \frac{32 \, c_3^2 \, g_S^2 \, m_{\dm}^4 \left(m^2_\med - 4 m^2_\dm\right)}
	{\pi v^2\left[\left(m_\med^2-4m_\dm^2\right)^2+m^2_\med\,\Gamma_\med^2\right]^2}.
\end{align} 
Quantitatively, we considered the $3\sigma$ range of the DM abundance as presented in the latest results of the Planck collaboration~\cite{Ade:2015xua}, $\Omega_{DM} h^2 = [0.1118,0.1199]$, assuming the $\Lambda$CDM model. As for the constraints from direct detection searches, the presence of a pseudoscalar mediator ensures that the implied nucleon scattering matrix element vanishes, in a way that the bound is simply evaded.

In the reported figures, the green squares highlight the points of the parameter space that satisfy both the LHC constraint (mono-jet excluded) and the DM abundance bound detailed above. 

\subsection{Fermi LAT constraints}

As for the constraints brought by the Fermi LAT measurements in indirect DM searches, the model predicts two distinct $\gamma$-ray signatures: 
\begin{enumerate}[i)]
	\item the dominant DM annihilation channel $SS \to gg$ gives rise to a distributed $\gamma$-ray signal consequently to the hadronization of the primary gluons. The corresponding signal shapes are presented in~\cite{Cirelli:2010xx}.
	\item the sub-dominant $SS \to \gamma \gamma$ annihilation channel, necessarily implied by the LHC signal, results in a $\gamma$-ray line at the mass of the DM candidate $S$ \cite{Huang:2015svl}.
\end{enumerate}

The ratio of the annihilation rates for the mentioned channels is 
$\langle \sigma v \rangle_{SS \to gg}/ \langle \sigma v \rangle_{SS \to \gamma\gamma} = \Gamma_{gg}/\Gamma_{\gamma\gamma} = 8 c_3^2/(c_1 \cos \theta_W + c_2 \sin \theta_W)^2$, which typically does not exceed $10^{-2}$ on the considered parameter ranges. We remark that, despite this suppression, a sharp $\gamma$-line in the quoted [30, 100] GeV range is far more distinguishable from the astrophysical power-law-like backgrounds than the milder signal brought by the $gg$ annihilation channel. Thus, in principle, both these channels can play a relevant role in indirect tests of the presented scenario. 

\begin{figure}[h]
\centering
\includegraphics[width=0.75\textwidth]{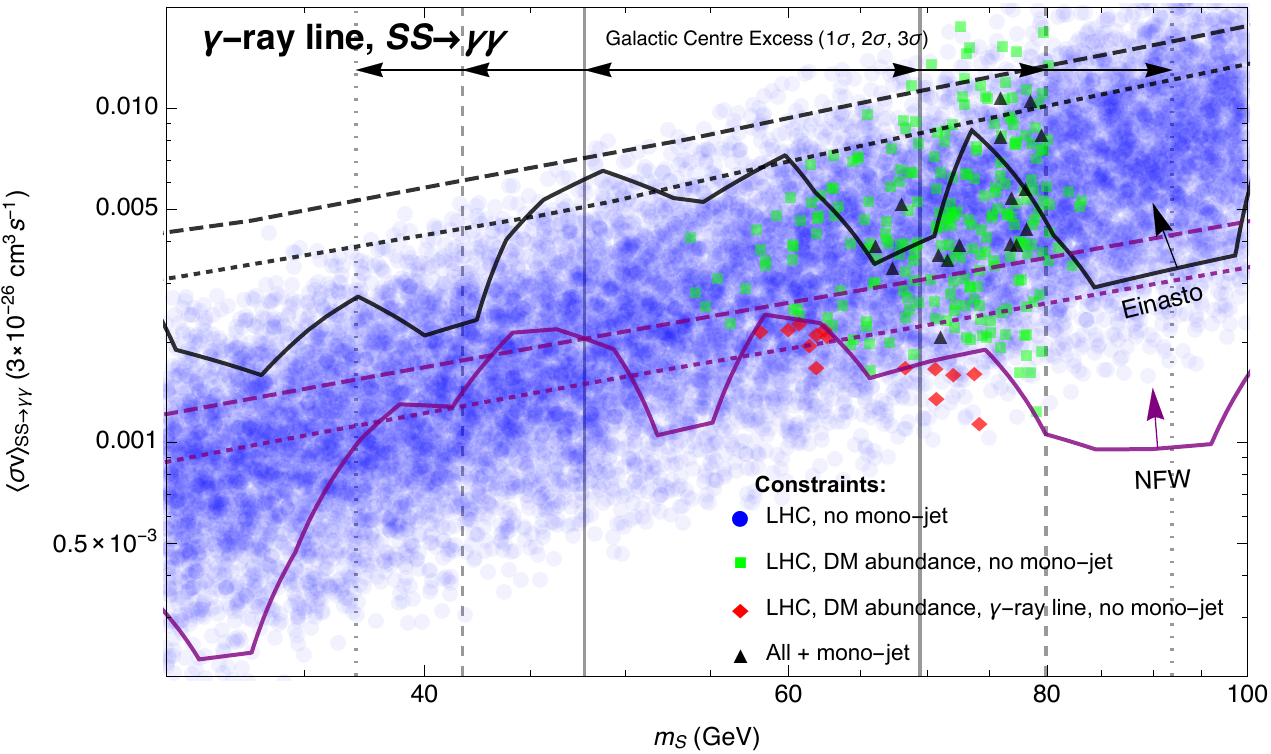}
\caption{Scatter plot for the DM mass $m_{S}$ and averaged annihilation cross-section times speed of the $SS \to \gamma\gamma$ channel. The colour code of the points is the same as in Fig.~\ref{fig:scatterplots}.  The black (purple) solid line denotes the Fermi LAT constraint for the Einasto (constrained NFW) profile. The solid curves show the observed limits, the dotted lines correspond to the 68\% contamination and the dashed ones to the 95\% contamination limits (see~\cite{Ackermann:2015lka} for details).}\label{fig:GRL}
\end{figure}

We show in Fig.~\ref{fig:GRL} the $\gamma$-ray line constraints derived from  the Fermi LAT data~\cite{Ackermann:2015lka}. The solid curves denote the experimental constraints obtained for the Einasto (black) and NFW (red) profiles. The dotted (dashed) lines illustrate instead the corresponding limits for the expected 68\% (95\%) containment (see~\cite{Ackermann:2015lka}). The red diamonds denote points that reproduce a correct DM abundance and satisfy the strictest Fermi LAT constraint, obtained with the NFW profile, but interestingly fail to observe the LHC mono-jet bound. The points that respect the latter, represented again by black triangles, are only compatible with the observed exclusion once the Einasto profile is assumed. Even in this case, however, the corresponding points are all in line only with the weakest of the possible constraints, corresponding to the 95\% containment limit indicated in the plot by a dashed black line.

\subsection{Reproducing the Galactic Centre Excess}

\begin{figure}[h]
\centering
\includegraphics[width=0.75\textwidth]{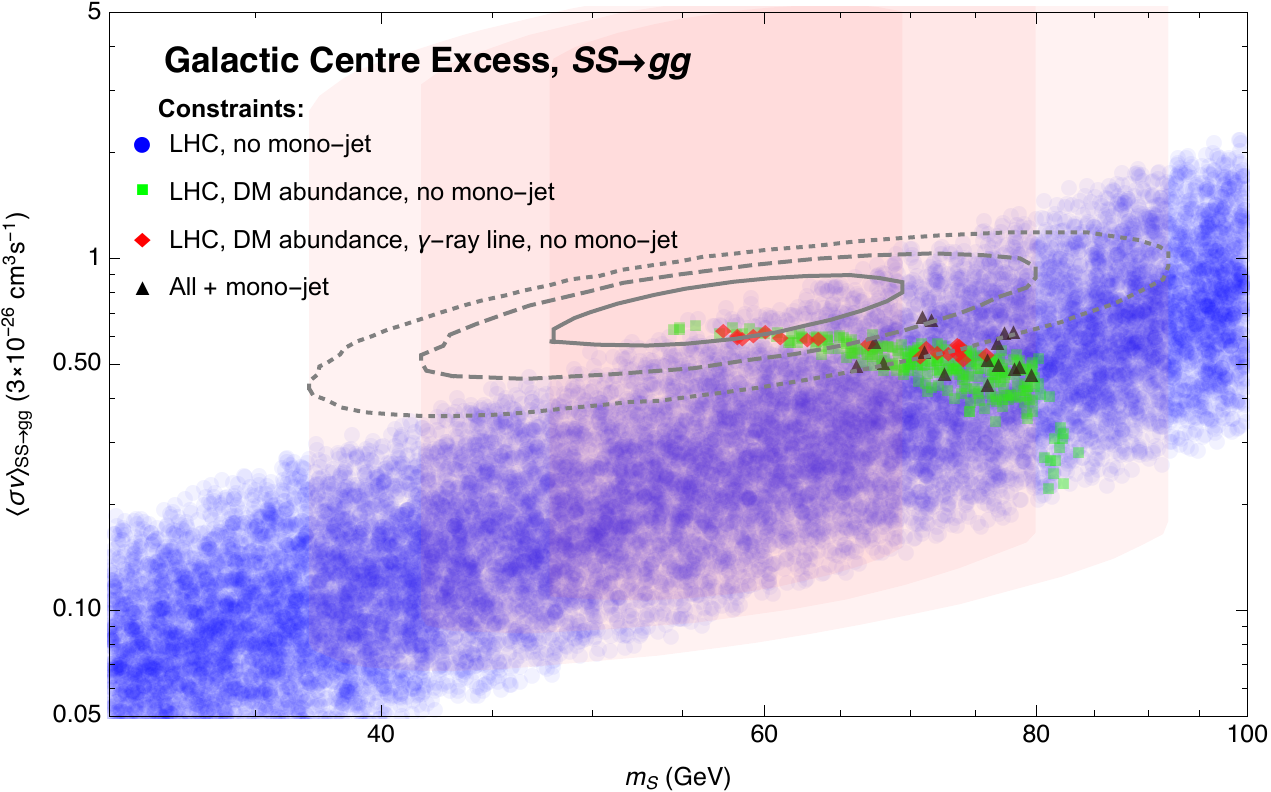}
\caption{Scatter plot for the DM mass, $m_S$, versus the averaged annihilation cross-section times velocity of $SS \to gg$. The colour code of the points is the same as in Fig.~\ref{fig:scatterplots}. The gray ovals denote the 1, 2 and 3$\sigma$ regions of the GCE fit for the NFW profile. The pink regions show the uncertainties on the fit due to possible different DM Galaxy profiles~\cite{Calore:2014nla}.}\label{fig:GCE}
\end{figure}

We now investigate whether the constraints analysed above allow for an explanation of the GCE signal within our framework. To this purpose, in Fig.~\ref{fig:GCE} we present a projection of the considered parameter space on the plane spanned by the DM mass, $m_S$, and the averaged annihilation cross section time velocity for the $gg$ channel, $\langle \sigma v \rangle_{SS \to gg}$. Intriguingly, we find that even after applying the constraints brought by LHC, the measurements of DM abundance as well as by the Fermi LAT $\gamma$-ray line searches, a large fraction of the generated points is well compatible with the GCE signal. We find, however, that whereas points satisfying the strictest Fermi LAT bound (illustrated by red diamonds in Fig.~\ref{fig:GCE}) are equally distributed within the $1\sigma$ and $2\sigma$ regions of the GCE fit, the points which also satisfy the mono-jet bound (black triangles) tend to accumulate in or towards the $2\sigma$ region. We argue that future improvements in the determination of the galactic dark matter profile, for instance by the GAIA mission, have therefore the potential to probe the proposed solution.  The $1\sigma$, $2\sigma$ and $3\sigma$ fitting regions of the GCE are here enclosed in the solid, dashed and dotted gray ovals respectively, for a considered NFW profile. The pink areas show the corresponding fit for other profiles and the uncertainty associated to the involved parameters; we refer to~\cite{Calore:2014nla} for more details. 
From Fig.~\ref{fig:GCE} it is also clear that reproducing the measured DM abundance (green squares, red diamonds and black triangles) strongly limits the parameter space yielding a viable GCE. Nevertheless, we remind that the constraint associated to DM abundance could be straightforwardly relaxed by considering, for instance, multicomponent DM or alternative DM production mechanism such as the DM freeze-in.

\FloatBarrier

\subsection{Cosmic antiprotons}

\begin{figure}[h]
\centering
\includegraphics[width=0.75\textwidth]{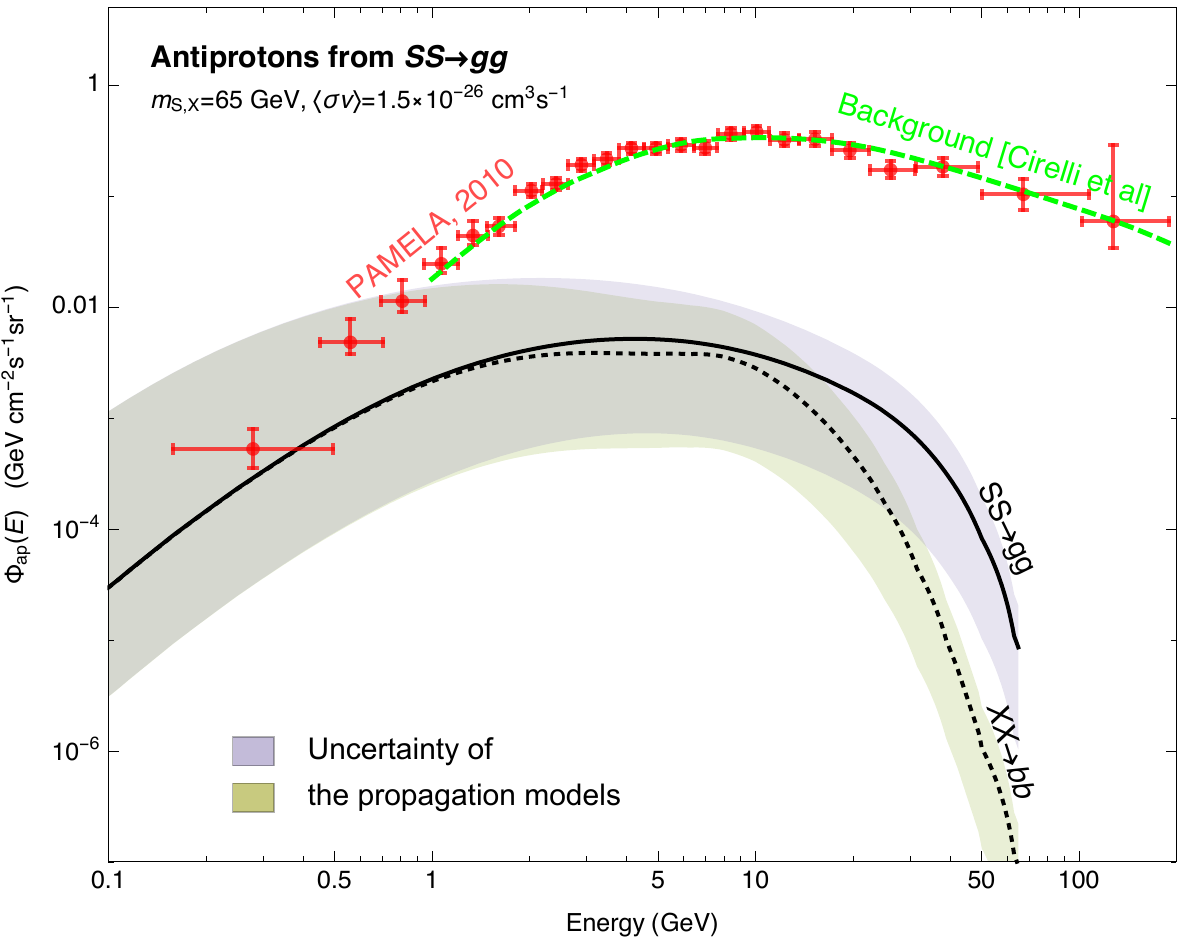}
\caption{The cosmic antiproton energy spectrum generated in DM annihilations compared to the astrophysical measured (red bars~\cite{Adriani:2010rc}) and modelled (dashed green~\cite{Cirelli:2013hv}) backgrounds. The antiproton spectrum resulting from DM annihilation is presented for $b\bar b$ and $gg$ final states respectively with a dotted and a solid  black line. The spectra are computed assuming the same DM mass and annihilation cross-section. The bands show the uncertainties due to cosmic ray propagation in the Galaxy~\cite{Cirelli:2010xx}.}\label{fig:antiprotons}
\end{figure}

The impact of cosmic antiproton measurements on DM phenomenology was previously demonstrated for instance by Bringmann et al ~\cite{Bringmann:2014lpa}, who employed the detected spectrum of cosmic ray antiprotons to forbid a part of the parameter space of DM candidates annihilating into $b \bar b$. Unfortunately, an analogous study for the $gg$ final state relevant to this work is not yet present in literature. Hence, to perform a first analysis, we  applied the PPPC4DMID toolkit~\cite{Cirelli:2010xx} to investigate the features of the antiproton spectrum yielded by the $gg$ final state. In Fig.~\ref{fig:antiprotons} we compare the antiproton spectra resulting from the DM annihilation into $b \bar b$ and $gg$ to the spectrum measured by the PAMELA experiment~\cite{Adriani:2010rc} and a phenomenological model proposed by Cirelli et al~\cite{Cirelli:2013hv}. As we can see, the $gg$ channel dominates by an order of magnitude the antiproton flux resulting from DM annihilations at the very high energy end of the spectrum. At low energies instead, $E \lesssim 20$~GeV, the spectra brought by $gg$ and  $b \bar b$ are essentially indistinguishable within the uncertainties of propagation models. 

As made clear by Fig.~\ref{fig:antiprotons}, the PAMELA constraints on the produced antiprotons affects mainly this low-energy tail of the spectrum, $E \lesssim 20$~GeV, essentially because the PAMELA measurement precision is at its best in the range $[2,20]$~GeV and because the DM annihilation antiproton spectra overtake the observed and modelled ones only for $E<0.2$~GeV. Unfortunately, the uncertainties due to solar modulation and other effects of cosmic ray propagation are very large for $E \lesssim 1$~GeV. Hence, we conclude that the PAMELA dataset constrains our scenario in the same measure as in the $b \bar b$ case studied in~\cite{Bringmann:2014lpa}, yielding no further bound on the regions of the parameter space already selected by the LHC and Fermi LAT constraints. Nevertheless, we point out that the future AMS-02 data release~\cite{AMS02:antiprotons} and a better modelling of solar modulation effects have clearly the potential to probe the presented $gg$ annihilation signature, to an extent that will be quantified in a dedicated work.

\FloatBarrier

\section{Conclusions}
\label{sec:Conclusions}

In this paper we investigated the compatibility between the LHC di-photon excess and the photon excess at the Galactic Centre by means of effective field theory methods. In our setup, a 750 GeV pseudoscalar particle connects the visible and dark sector of the Universe mediating the interaction that result in the mentioned excesses. Dark matter is here modelled after a scalar particle stabilised by a $Z_2$ symmetry. In our analysis we organically considered the following constraints
\begin{itemize}
	\item \textbf{LHC searches:} di-jet, di-boson and mono-jet constraints
	\item \textbf{DM relic abundance:} as measured by the Planck collaboration
	\item \textbf{Indirect $\mathbf{\gamma}$-ray signatures:} as measure by the Fermi LAT experiment. 
\end{itemize} 
Performing a scan of the parameter space of the model, we found that the mono-jet and Fermi LAT constraints strongly restrict the region where the di-photon excess and the Galactic Centre excess can be simultaneously reproduced. The surviving points are well in agreement with the collider bounds, although the mono-jet searches tend to favor a region of the parameter space towards the $2\sigma$ confidence interval of the Galactic Centre excess fit. Future improvements in the determination of the galactic dark matter profile have therefore the potential to probe the scenario. 
We have also analysed the implications of our scenario on the production of cosmic antiprotons via dark matter annihilation, finding that future release of antiproton flux data and improved modelling of solar modulation effects might provide further bounds.
     
\acknowledgments

The authors acknowledge the Estonian Research Council for supporting their work with the grants PUT808 and PUTJD110. AH thanks the Horizon 2020 programme as this project has received funding from the European Union’s Horizon 2020 research and innovation programme under the Marie Sklodowska-Curie grant agreement No 661103. This work was supported also by the grants IUT23-6, CERN+, due to the European Social Fund, and by the EU through the ERDF CoE program.

\bibliographystyle{JHEP}
\bibliography{2HDMS.bib}

\end{document}